\documentclass[10pt,letterpaper]{article}\usepackage[]{graphicx}\usepackage[]{color}
\pdfoutput=1

\makeatletter
\def\maxwidth{ %
  \ifdim\Gin@nat@width>\linewidth
    \linewidth
  \else
    \Gin@nat@width
  \fi
}
\makeatother

\definecolor{fgcolor}{rgb}{0.345, 0.345, 0.345}

\usepackage{framed}
\makeatletter
 {\par\unskip\endMakeFramed%
 \at@end@of@kframe}
\makeatother

\definecolor{shadecolor}{rgb}{.97, .97, .97}
\definecolor{messagecolor}{rgb}{0, 0, 0}
\definecolor{warningcolor}{rgb}{1, 0, 1}
\definecolor{errorcolor}{rgb}{1, 0, 0}
\newenvironment{knitrout}{}{} 

\usepackage{alltt}

\usepackage{cogsci}
\usepackage{pslatex}
\usepackage{apacite}
\usepackage{amsmath}
\usepackage{amsfonts}
\usepackage{verbatim}

\usepackage{tikz}
\usetikzlibrary{calc}
\newcommand\tikzmark[1]{\tikz[remember picture, baseline=(#1.base)] \node[anchor=base,inner sep=0pt, outer sep=0pt] (#1) {#1};}

\tikzset{
    ncbar angle/.initial=90,
    ncbar/.style={
        to path=(\tikztostart)
        -- ($(\tikztostart)!#1!\pgfkeysvalueof{/tikz/ncbar angle}:(\tikztotarget)$)
        -- ($(\tikztotarget)!($(\tikztostart)!#1!\pgfkeysvalueof{/tikz/ncbar angle}:(\tikztotarget)$)!\pgfkeysvalueof{/tikz/ncbar angle}:(\tikztostart)$)
        -- (\tikztotarget)
    },
    ncbar/.default=0.5cm,
}

\newcommand{\arrow}[2]{\begin{tikzpicture}[remember picture,overlay]
\draw[->,shorten >=3pt,shorten <=3pt] (#1.base) to [ncbar=\arrowht] (#2.base);
\end{tikzpicture}
\setlength{\arrowht}{0ex}
}
\usepackage{gb4e}

\newlength{\arrowht}
\renewcommand\glt{\vskip -\topsep}

\newcommand\arrowex{\setlength{\arrowht}{2.5ex}\ex}
\usepackage{tabulary}

\title{Modelling dependency completion in sentence comprehension \\as a Bayesian hierarchical mixture process: \\
A case study involving Chinese relative clauses}

 \author{{\large \bf Shravan Vasishth (vasishth@uni-potsdam.de)} \\
 Department of Linguistics, University of Potsdam, Germany.\\
  \AND {\large \bf Nicolas Chopin (nicolas.chopin@ensae.fr)} \\
  \'Ecole Nationale de la Statistique et de l'administration \'economique, Malakoff, France.\\
  \AND {\large \bf Robin Ryder (ryder@ceremade.dauphine.fr)} \\
Centre de Recherche en Math\'ematiques de la D\'ecision, CNRS, UMR 7534, Universit\'e Paris-Dauphine,\\ PSL Research University, Paris, France.\\
  \AND {\large \bf Bruno Nicenboim (bruno.nicenboim@uni-potsdam.de)} \\
 Department of Linguistics, University of Potsdam, Germany.\\  
  }
\IfFileExists{upquote.sty}{\usepackage{upquote}}{}
\IfFileExists{upquote.sty}{\usepackage{upquote}}{}
\begin{document}

\maketitle

\begin{abstract}
We present a case-study demonstrating the usefulness of Bayesian hierarchical mixture modelling for investigating cognitive processes.  
In sentence comprehension, it is widely assumed that the distance between linguistic co-dependents affects the latency of dependency resolution: the longer the distance, the longer the retrieval time (the distance-based account). An alternative theory, direct-access, assumes that retrieval times are a mixture of two distributions: one distribution represents successful retrievals (these are independent of dependency distance) and the other represents an initial failure to retrieve the correct dependent, followed by a reanalysis that leads to successful retrieval. We implement both models as Bayesian hierarchical models and show that the direct-access model explains Chinese relative clause reading time data better than the distance account.

\textbf{Keywords:} 
Bayesian Hierarchical Finite Mixture Models; Psycholinguistics; Sentence Comprehension; Chinese Relative Clauses; Direct-Access Model; K-fold Cross-Validation
\end{abstract}

\section{Introduction}

Bayesian cognitive modelling \cite{lee2014bayesian}, using probabilistic programming languages like JAGS \cite{plummer2011jags}, is an important tool in cognitive science. We present a case study from sentence processing research showing how hierarchical mixture models can be profitably used to develop probabilistic models of cognitive processes. Although the case study concerns a specialized topic in psycholinguistics, 
the approach developed here will be of general interest to the cognitive science community. 

In sentence comprehension research, dependency completion is assumed by many theories to be a key event. For example, consider a sentence such as (\ref{ex:themanslept}):

\begin{exe}
\ex  \label{ex:themanslept}
\begin{xlist}
\arrowex 
\tikzmark{The man} (on the bench) \tikzmark{was sleeping}
\arrow{was sleeping}{The man}
\end{xlist}
\end{exe}

\noindent
In order to understand who was doing what, the noun \textit{The man} must be recognized to be the subject of the verb phrase \textit{was sleeping}; this dependency is represented here as a directed arrow.
One well-known proposal \cite{jc92}, which we will call the \emph{distance account}, is that 
dependency distance between linguistically related elements partly determines comprehension difficulty as measured by reading times or question-response accuracy. 
For example, the Dependency Locality Theory (DLT) by \citeA{gibson00} and the cue-based retrieval account of \citeA{lewisvasishth:cogsci05} both assume that the longer the distance between two co-dependents such as a subject and a verb, the greater the retrieval difficulty at the moment of dependency completion. As shown in (\ref{ex:themanslept}), the distance between co-dependents can increase if a phrase intervenes.

As another example, consider the self-paced reading study in \citeA{gibsonwu} in Chinese subject and object relative clauses. The dependent variable here was the reading time at the head noun (\textit{official}). As shown in (\ref{ex:chineseRCs}), the distance between the head noun and the gap it is coindexed with is larger in subject relatives compared to object relatives.\footnote{The dependency could be equally well be between the relative clause verb and the head noun; nothing hinges on assuming a gap-head noun dependency.}
Thus, the distance account predicts an object relative advantage. For simplicity, we operationalize distance here as the number of words intervening between the gap inside the relative clause and the head noun. In the DLT, distance is operationalized as the number of  (new) discourse referents intervening between two co-dependents; and in the cue-based retrieval model, distance is operationalized in terms of decay in working memory (i.e., time passing by).

\begin{exe}
\ex  \label{ex:chineseRCs}
\begin{xlist}
\arrowex 
Subject relative
\gll [\tikzmark{GAP}$_i$ yaoqing fuhao de] \tikzmark{guanyuan}$_i$ xinhuaibugui \\
GAP invite tycoon DE official {have bad intentions}\\
\arrow{guanyuan}{GAP}
\glt `The official who invited the tycoon has bad intentions.’
\arrowex
Object relative 
\gll [fuhao yaoqing \tikzmark{GAP}$_i$ de] \tikzmark{guanyuan}$_i$ xinhuaibugui \\
tycoon invite GAP DE official { have bad intentions}\\
\arrow{guanyuan}{GAP}
\glt `The official who the tycoon invited has bad intentions.’
\end{xlist}
\end{exe}

In the Gibson and Wu study,
reading times were recorded using self-paced reading in the two conditions, with $37$ subjects and $15$ items, presented in a standard Latin square design. The experiment originally had $16$ items, but one item was removed in the published analysis due to a mistake in the item.
We coded subject relatives as $-1/2$, and object relatives as $+1/2$; this implies that an overall object relative advantage would show a negative coefficient. In other words, an object relative advantage corresponds to a negative sign on the estimate. 

The distance account's predictions can be evaluated by fitting the hierarchical linear model shown in (\ref{eq:lmm1}). Assume that (i)  $i$ indexes participants, $i=1,\dots,I$ and $j$ indexes items, $j=1,\dots,J$; (ii) $y_{ij}$ is the reading time in milliseconds for the $i$-th participant reading the $j$-th item;
 and (iii) the predictor $X$ is sum-coded ($\pm 1/2$), as explained above. Then,  the data $y_{ij}$ (reading times in milliseconds) are defined to be generated by the following model:

\begin{equation} \label{eq:lmm1}
y_{ij} = \beta_0+ \beta_1 X_{ij} + u_i + w_j + \varepsilon_{ij}
\end{equation}

\noindent
where $u_i \sim Normal(0,\sigma_u^2)$, $w_j \sim Normal(0,\sigma_w^2)$ and $\varepsilon_{ij} \sim Normal(0,\sigma_e^2)$; all three sources of variance are assumed to be independent. The terms $u_i$ and $w_j$ are called varying intercepts for participants and items respectively; they represent by-subject and by-item adjustments to the fixed-effect intercept $\beta_0$. 
Their variances, $\sigma_u^2$ and $\sigma_w^2$ represent between-participant (respectively item) variance.

This model is effectively a statement about the generative process that produced the data. If the distance account is correct, we would expect to find evidence that the slope $\beta_1$ is negative; specifically, reading times for object relatives are expected to be shorter than those for subject relatives. 
As shown in Table~\ref{tab:gworigresults}, this prediction appears, at first sight, to be borne out. Subject relatives are estimated to be read 120 ms slower than object relatives, apparently consistent with the predictions of the distance account.

\begin{table}[ht]
\centering
\begin{tabular}{rrrr}
  \hline
 & Estimate & Std. Error & t value \\ 
  \hline
$\hat\beta_0$ & 548.43 & 51.56 & 10.64* \\ 
$\hat\beta_1$ & -120.39 & 48.01 & -2.51* \\ 
   \hline
\end{tabular}
\caption{A linear mixed model using raw reading times in milliseconds as dependent variable, corresponding to the reported results in Gibson and Wu 2013. Statistical significance is shown by an asterisk.}\label{tab:gworigresults}
\end{table}

The object relative advantage shown in Table~\ref{tab:gworigresults} was originally presented in \citeA{gibsonwu} as a repeated measures ANOVA. 

To summarize, the conclusion from the above result would be that in Chinese, subject relatives are harder to process than object relatives because the gap inside the relative clause is more distant from the head noun in subject vs.\ object relatives. This makes it more difficult to complete the gap-head noun dependency in subject relatives. This distance-based explanation of processing difficulty is plausible given the considerable independent evidence from languages such as English, German, Hindi, Persian and Russian that dependency distance can affect reading time (see review in \citeA{SafaviEtAlFrontiers2016}). 

\begin{figure}[ht]
\begin{center}
\begin{knitrout}
\definecolor{shadecolor}{rgb}{0.969, 0.969, 0.969}\color{fgcolor}
\includegraphics[width=\maxwidth]{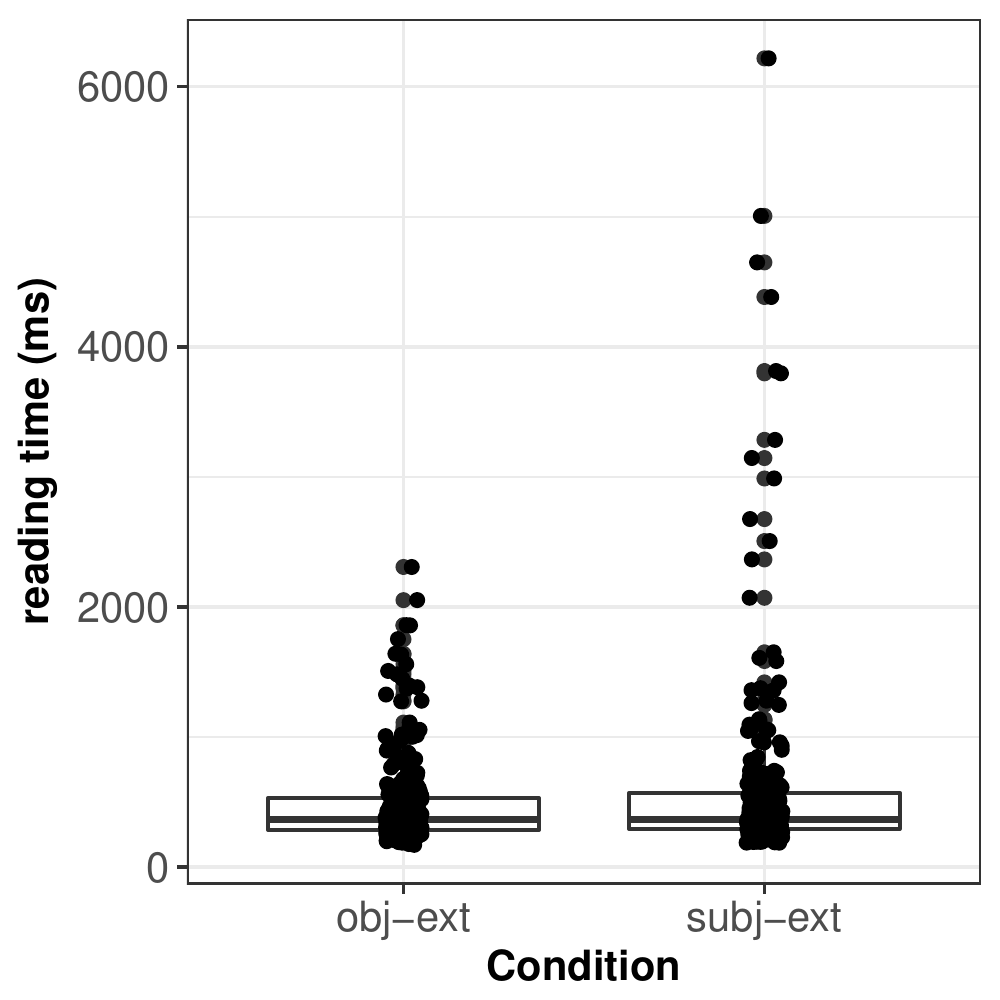} 

\end{knitrout}
\end{center}
\caption{Boxplots showing the distribution of reading times by condition of the \citeA{gibsonwu} data.} 
\label{fig:boxplots1}
\end{figure}

However, 
the distributions of the reading times for the two conditions show an interesting asymmetry that cannot be straightforwardly explained by the distance account. At the head noun, the reading times in subject relatives are much more spread out than in object relatives. This is shown in Figure~\ref{fig:boxplots1}, where reading times are shown on the log scale. Although this spread was ignored in the original analysis, a standard response to heterogeneous variances (heteroscedasticity) is to delete ``outliers'' based on some criterion; a common criterion is to delete all data lying beyond $\pm 2.5 SD$ in each condition.\footnote{In the published paper, \citeA{gibsonwu} did not delete any data, leading to the results shown in Table~\ref{tab:gworigresults}.}
This procedure assumes that the data points identified as extreme are irrelevant to the question being investigated. An alternative approach is to not delete data but to downweight the extreme values by applying a variance stabilizing transform \cite{box1964analysis}.  
Taking a log-transform of the reading time data, or a reciprocal transform, can reduce the heterogeneity in variance; see \citeA{VasishthetalPLoSOne2013} for analyses of the Gibson and Wu data using a transformation. 

One might think that if subject and object relatives are generated by LogNormal distributions with different means, then modelling the data as being generated by LogNormals would adequately explain the data.  
Table~\ref{tab:gworiglogresults} shows that if we assume such a model, there is no longer a statistically significant object relative advantage: the absolute t-value for the estimate of the $\beta_1$ parameter is smaller than the critical value of $2$ \cite{lme4new}. Thus, assuming that the data are generated by LogNormal distributions with different means for the subject and object relatives leads to the conclusion that there isn't much evidence for the distance account. 

\begin{table}[ht]
\centering
\begin{tabular}{rrrr}
  \hline
 & Estimate & Std.\ Error & t value \\ 
  \hline
$\hat\beta_0$ & 6.06  & 0.07 &   92.64*\\
$\hat\beta_1$ & -0.07 & 0.04 &   -1.61\\
   \hline
\end{tabular}
\caption{A linear mixed model using log reading times in milliseconds as dependent variable in the Gibson and Wu, 2013, data.}\label{tab:gworiglogresults}
\end{table}

Consider next the possibility that the heteroscedasticity in subject and object relatives in the Gibson and Wu data reflects a systematic difference in the underlying generative processes of reading times in the two relative clause types.
We investigate this question by modelling the extreme values as being generated from a mixture distribution.

Using the probabilistic programming language Stan \cite{stan-manual:2016}, we show
that a hierarchical mixture model provides a better fit to the data (in terms of predictive accuracy)  than several simpler hierarchical models.  As \citeA{NicenboimRetrieval2017} pointed out, the underlying generative process implied by a mixture model is consistent with the direct-access model of \citeA{mcelreeforakerdyer03}. We therefore suggest that, at least for the Chinese relative clause data considered here, the direct-access model may be a better way to characterize the dependency resolution process than the distance account.

We can implement the direct-access model as a hierarchical mixture model with retrieval time assumed to be generated from one of two distributions, where the proportion of trials in which a retrieval failure occurs (the mixing proportion) is $p_{sr}$ in subject relatives, and $p_{or}$ in object relatives. 
The expectation here is the extreme values that are seen in subject relatives are due to $p_{sr}$ being larger than  $p_{or}$.

\begin{equation} \label{eq:mixmod}
\begin{split} 
\hbox{Subject relatives} & \\
y_{ij} \sim& p_{sr} \cdot LogNormal(\beta+\delta+u_i+w_j,\sigma_{e'}^2) \\
  +& (1-p_{sr})\cdot LogNormal(\beta+u_i+w_j,\sigma_e^2)\\
\hbox{Object relatives} & \\
y_{ij} \sim& p_{or} \cdot LogNormal(\beta+\delta+u_i+w_j,\sigma_{e'}^2) \\
  +& (1-p_{or})\cdot LogNormal(\beta+u_i+w_j,\sigma_e^2)\\
\end{split}
\end{equation}

\noindent
Here, the terms $u_i$ and $w_j$ have the same interpretation as in equation~\ref{eq:lmm1}.


\subsection{Model comparison}

Bayesian model comparison can be carried out 
using different methods. Here, we use Bayesian k-fold cross-validation as discussed in \citeA{vehtari2016LOOwaic}. 
This method evaluates the predictive performance of alternative models, and models with different numbers of parameters can be compared \cite{vehtari2012survey,gelman2014understanding}. 

The k-fold cross-validation algorithm is as follows:

\begin{enumerate}
\item
Split data pseudo-randomly into $K$ \textit{held-out} sets $\mathbf{y}_{(k)}$, where $k=1,\dots,K$ that are a fraction of the original data, 
and $K$ \textit{training sets}, $\mathbf{y}_{(-k)}$. 
Here, we use $K=10$, and the length of the held-out data-vector $\mathbf{y}_{(k)}$ is approximately $1/K$-th the size of the full data-set. 
We ensure that each participant's data appears in the training set and contains an approximately balanced number of data points for each condition.
\item 
Sample from the model using each of the $K$ training sets, 
and obtain posterior distributions $p_{\hbox{post(-k)}} (\mathbf{\theta}) = p(\mathbf{\theta}\mid \mathbf{y}_{(-k)})$, where $\mathbf{\theta}$ is the vector of model parameters.  
\item 
Each posterior distribution $p(\theta\mid \mathbf{y}_{(-k)})$ is used to compute predictive accuracy for each held-out data-point $y_i$:

\begin{equation}
\log p(y_i \mid \mathbf{y}_{(-k)}) = \log \int p(y_i \mid \mathbf{\theta}) p(\mathbf{\theta}\mid \mathbf{y}_{(-k)})\, d\mathbf{\theta}
\end{equation}

\item 
Given that the posterior distribution $p(\mathbf{\theta}\mid \mathbf{y}_{(-k)})$ is summarized by $s=1,\dots,S$ simulations, i.e.,  $\mathbf{\theta}^{k,s}$, log predictive density for each data point $y_i$ in subset $k$ is computed as

\begin{equation}
\widehat{elpd}_i = \log \left(\frac{1}{S} \sum_{s=1}^S p(y_i\mid \mathbf{\theta}^{k,s})\right)
\end{equation}

\item 
Given that all the held-out data in the $K$ subsets are $y_i$, where 
$i=1,\dots,n$, we obtain the $\widehat{elpd}$ for all the held-out data points by summing up the $\widehat{elpd}_i$:

\begin{equation}
\widehat{elpd} = \sum_{i=1}^n \widehat{elpd}_i
\end{equation}

\end{enumerate}

The difference between the $\widehat{elpd}$'s of two competing models is a measure of relative predictive performance. We can also compute  the standard deviation of the sampling distribution (the standard error) of the difference in $\widehat{elpd}$ using the formula discussed in \citeA{vehtari2016LOOwaic}. Letting $\widehat{ELPD}$ be the vector  $\widehat{elpd}_1,\dots,\widehat{elpd}_n$, we can write: 

\begin{equation}
se(\widehat{elpd}_{m0} - \widehat{elpd}_{m1}) = 
\sqrt{n Var(\widehat{ELPD})}
\end{equation}

When we compare the model (\ref{eq:lmm1}) with (\ref{eq:mixmod}), if (\ref{eq:mixmod}) has a higher $\widehat{elpd}$, then it has a better predictive performance compared to (\ref{eq:lmm1}).  

The quantity $\widehat{elpd}$ is a Bayesian alternative to the Akaike Information Criterion \cite{akaike1974new}. Note that the relative complexity of the models to be compared is not relevant: the sole criterion here is out-of-sample predictive performance. As we discuss below (Results section), increasing complexity will not automatically lead to better predictive performance. See \citeA{vehtari2012survey,gelman2014understanding} for further details.\footnote{We also used a simpler method than k-fold cross-validation to compare the models; this method is described in \citeA{vehtari2016LOOwaic}. The results are the same regardless of the model comparison method used.}

\subsection{The data}

The evaluation of these models was carried out using two separate data-sets. The first was the original study from \citeA{gibsonwu} that was discussed in the introduction. The second study was a replication of the Gibson and Wu study that was published in \citeA{VasishthetalPLoSOne2013}. This second study served the purpose of validating whether independent evidence can be found for the mixture model selected using the original Gibson and Wu data.

\subsection{Results}

In the models presented below, the dependent variable
is reading time in milliseconds.
Priors are defined for the model parameters as follows. All standard deviations are constrained to be greater than 0 and have priors $\hbox{Cauchy}(0,2.5)$ \cite{Gelman14}; probabilities have priors $\hbox{Beta}(1,1)$; and all coefficients ($\beta$ parameters) have priors 
$\hbox{Cauchy}(0,2.5)$.

\paragraph{Fake-data simulation for validating model}
Before evaluating relative model fit, we first simulated data from a mixture distribution with known parameter values,  and then sampled from the models representing the distance account and the direct-access model. The goal of fake-data simulation was to validate the models and the model comparison method: with reference to the simulated data, we asked (a) whether the 95\% credible intervals of the posterior distributions of the parameters in the mixture model contain the true parameter values used to generate the data; and (b) whether k-fold cross validation can identify the mixture model as the correct one when the underlying generative process matches the mixture model. The answer to both questions was ``yes''. This raises our confidence that the models can identify the underlying parameters with real data. 
The fake-data simulation also showed that when the true underlying generative process was consistent with the distance account but not the direct access model, the hierarchical linear model and the mixture model had comparable predictive performance. In other words, the mixture model furnished a superior fit only when the true underlying generative process for the data was in fact a mixture process.
Further details are omitted here due to lack of space.

\subsubsection{The original Gibson and Wu study}

The estimates from the hierarchical linear model (equation~\ref{eq:lmm1}) and 
the mixture model (equation~\ref{eq:mixmod}) are shown in Tables~\ref{tab:lmm1results} 
and \ref{tab:mixmodresults}. Note that in Bayesian modelling we are not interested in ``statistical significance'' here; rather, the goal is inference and comparing predictive performance of two competing models. 

\begin{table}[!htbp]
\centering
\begin{tabular}{rrrr}
  \hline
 & mean & lower & upper \\ 
  \hline
$\hat\beta_1$ & 6.06 & 5.91 & 6.20 \\ 
  $\hat\beta_2$ & -0.07 & -0.16 & 0.02 \\ 
  $\hat\sigma_e$ & 0.52 & 0.49 & 0.55 \\ 
  $\hat\sigma_u$ & 0.25 & 0.18 & 0.34 \\ 
  $\hat\sigma_w$ & 0.20 & 0.12 & 0.33 \\ 
   \hline
\end{tabular}
\caption{Posterior parameter estimates from the hierarchical linear model (equation~\ref{eq:lmm1}) corresponding to the distance account. The data are from Gibson and Wu, 2013. Shown are the mean and 95\% credible intervals for each parameter.}\label{tab:lmm1results}
\end{table}

\begin{table}[!htbp]
\centering
\begin{tabular}{rrrr}
  \hline
 & mean & lower & upper \\ 
  \hline
$\hat\beta_0$ & 5.85 & 5.76 & 5.95 \\ 
  $\hat\delta$ & 0.93 & 0.73 & 1.14 \\ 
  $\hat{p}_{sr}-\hat{p}_{or}$ & 0.04 & -0.04 & 0.13 \\ 
  $\hat{p}_{sr}$ & 0.25 & 0.17 & 0.34 \\ 
  $\hat{p}_{or}$ & 0.21 & 0.14 & 0.29 \\ 
  $\hat\sigma_{e'}$ & 0.64 & 0.54 & 0.74 \\ 
  $\hat\sigma_e$ & 0.22 & 0.20 & 0.25 \\ 
  $\hat\sigma_u$ & 0.24 & 0.18 & 0.31 \\ 
  $\hat\sigma_w$ & 0.09 & 0.05 & 0.16 \\ 
   \hline
\end{tabular}
\caption{Posterior parameter estimates from the hierarchical mixture model (equation~\ref{eq:mixmod}) corresponding to the direct-access model. The data are from Gibson and Wu, 2013. Shown are the mean and 95\% credible intervals for each parameter.}\label{tab:mixmodresults}
\end{table}

Table~\ref{tab:mixmodresults} shows that the mean difference between the probability $p_{sr}$ and $p_{or}$ is 4\%; the posterior probability of this difference being greater than zero is 82\%.
K-fold cross-validation shows that $\widehat{elpd}$ for the hierarchical model is $-3761$ (SE: $38$) and for the mixture model 
is $-3614$ ($35$). The difference between the two $\widehat{elpd}$s is $148$ ($18$). The larger $\widehat{elpd}$ in the hierarchical mixture model suggests that it has better predictive performance than the hierarchical linear model. In other words, the direct-access model has  better predictive performance than the distance model.

\subsubsection{The replication of the Gibson and Wu study}



\begin{figure}[ht]
\begin{center}
\begin{knitrout}
\definecolor{shadecolor}{rgb}{0.969, 0.969, 0.969}\color{fgcolor}
\includegraphics[width=\maxwidth]{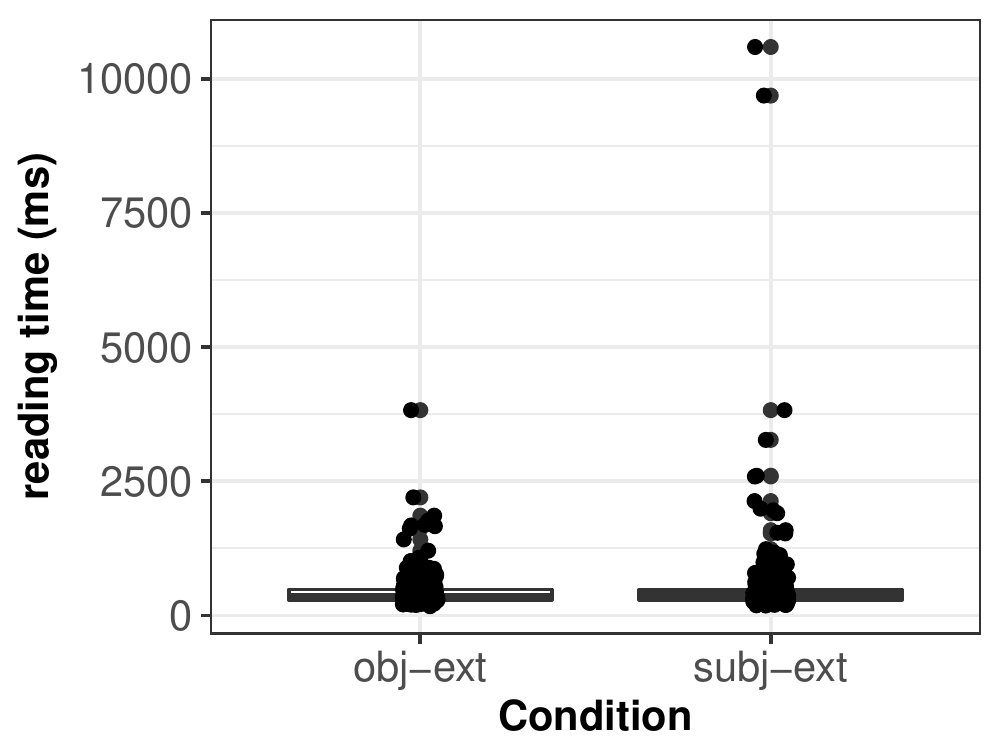} 

\end{knitrout}
\end{center}
\caption{Boxplots showing the distribution of reading times by condition of the replication of the Gibson and Wu data.} 
\label{fig:boxplots2}
\end{figure}

This data-set, originally reported by \citeA{VasishthetalPLoSOne2013}, had 40 participants and the same 15 items as in Gibson and Wu's data. Figure~\ref{fig:boxplots2} shows the distribution of the data by condition; there seems to a similar skew as in the original study, although the spread is not as dramatic as in the original study.

Tables~\ref{tab:lmm1resultsrep} and \ref{tab:mixmodresultsrep} show the estimates of the posterior distributions from the two models. 
Table~\ref{tab:mixmodresults} shows that the mean difference between the probability $p_{sr}$ and $p_{or}$ is 7\%; the posterior probability of this difference being greater than zero is 96\%.

The $\widehat{elpd}$ for the hierarchical model is $-3959$ ($53$), and for the hierarchical mixture model, $-3801$ ($38$). The difference in $\widehat{elpd}$ is $158$ ($29$). Thus, in the replication data as well, the predictive performance of the mixture model is better than the hierarchical linear model.

\begin{table}[!htbp]
\centering
\begin{tabular}{rrrr}
  \hline
 & mean & lower & upper \\ 
  \hline
$\hat\beta_0$ & 6.00 & 5.88 & 6.12 \\ 
$\hat\beta_1$ & -0.09 & -0.16 & -0.01 \\ 
$\hat\sigma_e$ & 0.44 & 0.41 & 0.47 \\ 
$\hat\sigma_u$ & 0.25 & 0.19 & 0.33 \\ 
$\hat\sigma_w$ & 0.16 & 0.10 & 0.26 \\ 
   \hline
\end{tabular}
\caption{Posterior parameter estimates from the hierarchical linear model (equation~\ref{eq:lmm1}) corresponding to the distance account. The data are from the replication of Gibson and Wu, 2013 reported in Vasishth et al., 2013. Shown are the mean and 95\% credible intervals for each parameter.}\label{tab:lmm1resultsrep}
\end{table}

\begin{table}[!htbp]
\centering
\begin{tabular}{rrrr}
  \hline
 & mean & lower & upper \\ 
  \hline
$\hat\beta_0$ & 5.86 & 5.78 & 5.95 \\ 
$\hat\delta$ & 0.75 & 0.56 & 0.97 \\ 
$\hat{p}_{sr}-\hat{p}_{or}$ & 0.07 & -0.01 & 0.15 \\ 
$\hat{p}_{sr}$ & 0.23 & 0.15 & 0.33 \\ 
$\hat{p}_{or}$ & 0.16 & 0.09 & 0.25 \\ 
$\hat\sigma_{e'}$ & 0.69 & 0.59 & 0.81 \\ 
$\hat\sigma_e$ & 0.21 & 0.18 & 0.23 \\ 
$\hat\sigma_u$ & 0.22 & 0.17 & 0.29 \\ 
$\hat\sigma_w$ & 0.07 & 0.04 & 0.12 \\ 
   \hline
\end{tabular}
\caption{Posterior parameter estimates from the hierarchical linear model (equation~\ref{eq:mixmod}) corresponding to the direct-access model. The data are from the replication of Gibson and Wu, 2013 reported in Vasishth et al., 2013. Shown are the mean and 95\% credible intervals for each parameter.}\label{tab:mixmodresultsrep}
\end{table}

\subsection{Discussion}

The model comparison and parameter estimates presented above suggest that, at least as far as the Chinese relative clause data are concerned, a better way to characterize the dependency completion process is in terms of the direct-access model and not the distance account implied by \citeA{gibsonwu} and \citeA{lewisvasishth:cogsci05}. Specifically, there is suggestive evidence in the \citeA{gibsonwu} data that a higher proportion of retrieval failures occurred in subject relatives compared to object relatives. In other words, increased dependency distance may have the effect that it increases the proportion of retrieval failures (followed by reanalysis).\footnote{A reviewer suggests that the direct-access model may simply be an elaboration of the distance model. This is by definition not the case: direct access (i.e., distance-independent access) is incompatible with the distance account.}

There is one potential objection to the conclusion above. It would be important to obtain independent evidence as to which dependency was eventually created in each trial. 
This could be achieved by asking participants multiple-choice questions to find out which dependency they built in each trial. Although such data is not available for the present study, in other work (on number interference) \cite{NicenboimEtAl2016} did collect this information. There, too, we found that the direct-access model best explains the data \cite{NicenboimRetrieval2017}. In future work on Chinese relatives, it would be helpful to carry out a similar study to determine which dependency was completed in each trial.
In the present work, the modelling at least shows how the extreme values in subject relatives can be accounted for by assuming a two-mixture process.





\section{Conclusion}

The mixture models suggest that, in the specific case of Chinese relative clauses, increased processing difficulty in subject relatives is not due to dependency distance leading to longer reading times, as suggested by \citeA{gibsonwu}. Rather, a more plausible explanation for these data is in terms of the direct-access model of \citeA{mcelreeforakerdyer03}. Under this view, retrieval times are not affected by the distance between co-dependents, but a higher proportion of retrieval failures occur in subject relatives compared to object relatives. This leads to a mixture distribution in both subject and object relatives, but the proportion of the failure distribution is higher in subject relatives. 

In conclusion, this paper serves as a case study demonstrating the flexibility of Bayesian cognitive modelling using finite mixture models. This kind of modelling approach can be used flexibly in many different research problems in cognitive science. One example is the above-mentioned work by \citeA{NicenboimRetrieval2017}. Another example, also from sentence comprehension, is the evidence for feature overwriting \cite{nairne1990feature} in parsing
\cite{VasishthEtAlICCM2017}.

\section{Acknowledgments}

We are very grateful to Ted Gibson for generously providing the raw data and the experimental items from \citeA{gibsonwu}. Thanks also go to Lena J\"ager for many insightful comments. Helpful observations by Aki Vehtari are also gratefully acknowledged.
For partial support of this research, 
we thank the Volkswagen Foundation through grant 89 953.

\bibliographystyle{apacite}

\setlength{\bibleftmargin}{.125in}
\setlength{\bibindent}{-\bibleftmargin}

\bibliography{/Users/shravanvasishth/Dropbox/Bibliography/bibcleaned.bib}

\end{document}